\documentstyle[12pt,epsfig]{article}                            

\newcommand{\Ftwo}   {\mbox{$F_2$}}
\newcommand{\Fz}   {\mbox{$F_3$}}
\newcommand{\FL}   {\mbox{$F_{_{L}}$}}

\newcommand{\QQ}  {\mbox{${Q^2}$}}

\newcommand{\sss}{\Sigma_e}
\newcommand{\SSS}{\Sigma}
\newcommand{\ggg}{\gamma}
\newcommand{\gss}{\gamma_{\Sigma}}
\newcommand{\beq}{\begin{equation}} 
\newcommand{\eeq}{\end{equation}}   

\newcommand{\dth}{{\delta \theta}}
\newcommand{\dxx}{{\large $\frac{\delta x}{x}$}}
\newcommand{\dyy}{{\large $\frac{\delta y}{y}$}}
\newcommand{\dqq}{{\large $\frac{\delta Q^2}{Q^2}$}}
\newcommand{\ddxx}{\frac{\delta x}{x}}
\newcommand{\ddyy}{\frac{\delta y}{y}}
\newcommand{\ddqq}{\frac{\delta Q^2}{Q^2}}
\newcommand{\daxx}{\frac{\delta x_{DA}}{x_{DA}}}
\newcommand{\dayy}{\frac{\delta y_{DA}}{y_{DA}}}
\newcommand{\daqq}{\frac{\delta Q^2_{DA}}{Q^2_{DA}}}
\newcommand{\dye}{\frac{\delta y_e}{y_e}}
\newcommand{\dxe}{\frac{\delta x_e}{x_e}}
\newcommand{\dqe}{\frac{\delta Q^2_e}{Q^2_e}}

\newcommand{\dee}{{$\frac{\delta E}{E}$}}
\newcommand{\dss}{{\large $\frac{\delta \SSS}{\SSS}$}}

\newcommand{\DD}{{\pm \Delta}}
\newcommand{\htab}{\rule[-3mm]{0mm}{10mm}}
\newcommand{\hhtab}{\rule[-5mm]{0mm}{12mm}}

\newcommand{\dtt}{{\large $\frac{\delta p_{T,h}}{p_{T,h}}$}}

\newlength{\dinwidth}       
\newlength{\dinmargin}      
\begin{document}   

\begin{titlepage}  

\begin{flushleft}  
\noindent  
{\tt DESY 97-137   } \\       
{\tt December 1997}       \\
\end{flushleft}    
\vspace*{3.0cm}    
\begin{center}     
\begin{Large} {\bf 
Structure Function Measurements and \\
Kinematic Reconstruction at HERA

%%  Kinematic Constraints at Low and High $Q^2$ \\
%%   in  Deep   Inelastic  Scattering at HERA     
%       
}\end{Large} \end{center}   
\vspace{2.0cm}     
\begin{center}     
\begin{large}      
Ursula Bassler, Gregorio Bernardi  \\       
\end{large}
\end{center}       
%vspace{0.8cm}     
%vspace{0.8cm}     
\begin{center}     
     Laboratoire de Physique Nucl\'eaire et des Hautes Energies\\
     Universit\'e Paris 6-7, 4 Place Jussieu,    
75252 Paris, France\\
{\it e-mail: bassler@mail.desy.de; gregorio@mail.desy.de}     
\end{center}       
\vspace{2.5cm}     
\begin{abstract}   
%noindent  
The procedure used for structure function measurements at HERA is briefly
described and related to the properties of kinematic reconstruction.
The  reconstruction methods of the inclusive deep 
inelastic scattering  variables are reviewed and their sensitivity to
the energy and angle miscalibrations are discussed in detail. New 
prescriptions are introduced and related to the standard methods
in order to optimize the $F_2$ structure function measurement over the widest
kinematic range, both in the low $x$, low $Q^2$ and in the high $x$,
high $Q^2$ regions. The prospects for the future high $Q^2$ studies
are briefly discussed.
\end{abstract}    
\end{titlepage}

\section{Introduction}      
The measurement of the structure functions of the nucleon 
is a major tool for the study of the strong interaction 
and the behaviour of the parton densities in the hadrons. To reveal 
possible non-standard small deviations
 from their well established behaviour described
by the DGLAP~\cite{bb.dglap} evolution equations,
such as BFKL effects at low $x$ or the 
presence of intrinsic charm in the proton at high $x$ to name just two of them,
requires  a precise reconstruction of the deep inelastic scattering (DIS) 
kinematics over the widest possible kinematic range. With the advent of      
the HERA  electron-proton collider, this reconstruction
no longer needs to rely      
on  the scattered lepton only, since the 
most important  part of the 
hadronic system is visible in the almost hermetic  H1 and     
ZEUS detectors. This redundancy  allows for   an
experimental control of the systematic 
errors and of the radiative corrections to the structure function    
measurement  if it is based on      
several independent methods to determine the  usual DIS
kinematic variables $x, y$, $Q^2$:
\begin{equation}   
  x = \frac{Q^2} {2 P.q}    \hspace*{1cm} y = \frac{P.q} {P.k}
 \hspace*{1cm}   
 Q^2= -(k-k')^2 = -q^2  = xys    
\end{equation}     
with $s$ being the $ep$ center of mass energy squared,     
 $P,k $  the 4-vectors of the incident proton and lepton,
and $k'$ of the scattered lepton.  
Since many different reconstruction methods have already been 
used at HERA,
it is a natural objective to optimize this reconstruction,
and to try to find ``the best'' method, or at least to justify the
use of a given method instead of a kinematic fitting procedure for
instance.
In this report we  briefly sketch in section 2 the  procedure 
used so far to measure $F_2$ at HERA, and relate it to 
the effects of kinematic reconstruction.
In section 3 we review the  methods of kinematic 
reconstruction used at HERA,  and we classify them 
on the basis of their properties. 
In section 4 we discuss possible improvements
of these reconstruction methods, in particular those 
related to the study of
low $x$ physics. 
In section 5 we study the effect of the hadronic final state
and  scattered electron  reconstruction errors on the kinematic 
methods in order to understand the choices made in this field
by the two HERA collaborations.  
In section 6,  the high $Q^2$ case is treated in more detail 
due to its future importance and also
since there are new possibilities in this kinematic regime.
In conclusion, we briefly provide some prospects on the influence of these
technical matters on the  structure function program of
the next decade.
%The specific reconstruction of very high $Q^2$ events is described 
%in~\cite{bb2}.

\section{Structure Function Measurements}     

The cross-section 
%\cite{csectDIS} 
for the DIS reaction 
 $e^{+}+p\rightarrow e^{+}+X$
with unpolarized beams is:
\begin{equation}
\frac{d^2\sigma}{dx\ dQ^2} = \frac{2\pi \alpha^2}{xQ^4}
\left[Y_+ \Ftwo (x,\QQ )-y^2\FL (x,\QQ )-Y_{-}x\Fz(x,\QQ )\right]
(1+\delta_r(x,\QQ ))
\end{equation}
In this equation $\alpha $ is the electromagnetic coupling,
\Ftwo\  is the generalized structure function which
reflects both photon and $Z^{\circ}$ exchange,
\FL\ is the longitudinal structure 
function, \Fz\ is the parity violating 
term arising only from the $Z^\circ$ exchange,
$\delta_r$ is the electroweak radiative correction.
The helicity dependence of electroweak interactions, is contained in the
functions 
$
Y_{\pm }(y)=1\pm (1-y)^2. 
$

At HERA, these three structure functions  of the proton can 
be measured, in particular $F_2$ which has already been measured
over several orders of magnitude in $x$ and $Q^2$ with a precision
of about 5 \%.
Although apparently only one observable ($\frac{d^2\sigma}{dx\ dQ^2}$)
is related to the three structure functions, 
the problem can be solved because in some
kinematic region only one of them has a relevant contribution and/or because
the beam conditions (energy, lepton sign) can be changed, hence changing
the  coefficient in front of each of  them, for a given  
$\frac{d^2\sigma}{dx\ dQ^2}$.
In the following, we will not study the procedure to derive them
from the cross-section measurement (which are described
in \cite{ingel} for instance), but concentrate on the 
determination of  $\frac{d^2\sigma}{dx\ dQ^2}$.

Generally, the value  
of  $\frac{d^2\sigma}{dx\ dQ^2}$ at the point ($x_o,Q^2_o$)
is experimentally determined by the number $N_{\Delta}^D$ 
({\small ``$D$''} stands for Data) of DIS events observed
in an ($x,Q^2$) bin $\Delta$ centered around  ($x_o,Q^2_o$), normalized
to the integrated luminosity $L^D$ accumulated during the data taking 
and corrected by an acceptance 
factor $T_{\Delta}^D$ which depends on  the
event selection cuts and on  the detector response:
\begin{equation}
\frac{d^2\sigma}{dx\ dQ^2} 
= \frac{{\it C}_\Delta}{L^D} \  \frac{N_{\Delta}^D} { T_{\Delta}^D}
\end{equation}
{\it C}$_\Delta$
 is a numerical factor which takes into account the surface of the
bin. Actually, resolution effects cause migration of events from one bin
to other bins, so the previous relation should be treated in 
fact as a matrix relation,
with the cross-section being  related to all bins 
via the inverse of a matrix  $T(ij)$,
in which every matrix element $(ij)$ gives the probability that an event 
originating from the bin $i$ is reconstructed in a bin $j$. Thus the double
differential cross-section measured at the center of the bin $(i)$ is:
\begin{equation}
\frac{d^2\sigma}{dx\ dQ^2} =  \frac{{\it C}_\Delta}{L^D} \ 
\sum_j {{N^D_j}  T^{-1}(ij)}
\end{equation}
The  matrix $T$ can be obtained by simulating a large sample of
DIS events, provided 
the simulation is able to reproduce the detector effects.
However to solve equation (4) one needs to invert this matrix.
This is a  non-trivial task due in particular to 
numerical instabilities~\cite{blobel,virchaux,zech}, 
which has not been done yet
for the published  HERA structure function measurements. 
Instead, an iterative procedure is used to solve equation (3).
Obviously, if  the structure function used  in the simulation
is equal to the one to be measured, and
assuming that  $T_{\Delta}^D =  T_{\Delta}^S$ ({\small ``$S$''}  stands
for simulation), we have:
\begin{equation}
\frac{d^2\sigma}{dx\ dQ^2}  
%
%= \frac{N_{\Delta}^D} {L^D T_{\Delta}^D}
%  \  \frac {L^S T_{\Delta}^S}{N_{\Delta}^S}
%  \frac{d^2\sigma^S}{dx\ dQ^2}
%
= \frac{N_{\Delta}^D} {L^D}
\  \frac {L^S}{N_{\Delta}^S}\frac{d^2\sigma^S}{dx\ dQ^2} 
\end{equation}
This equation is  solved iteratively,  starting  the simulation
from a ``guessed'' structure
function parametrization, and replacing it at each iteration by the
parametrization obtained from a fit to the structure function
 ``measured'' at the previous iteration.
In three or four iterations the result is stable within one \% or less
(see ~\cite{ZEUSF294} for a more detailed discussion) in the 
measureable regions.
 These measureable regions are empirically 
characterized by values of  $T_{\Delta}$ close to unity. In order
to define them in a more rigorous way and
to understand their relation to the kinematic reconstruction,
 we will now study  the  $T_{\Delta}$ factors.

\begin{figure}[htb]                                                           
\begin{center}                                
\epsfig{file=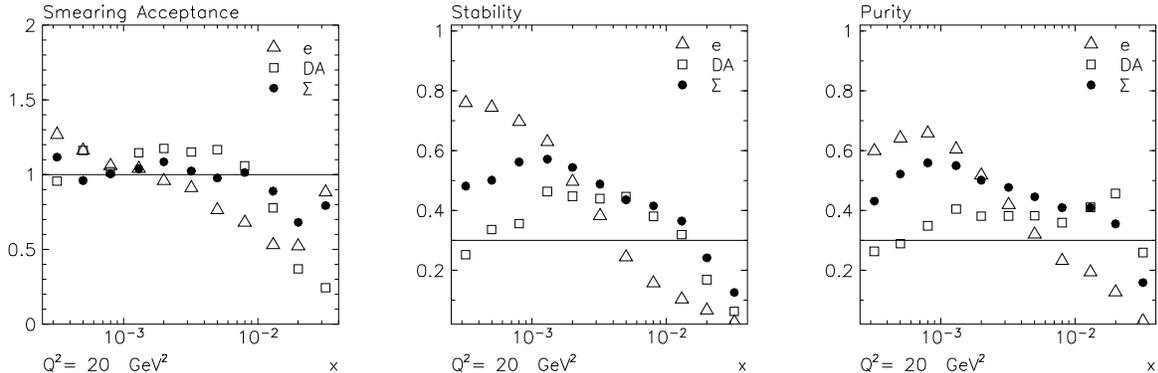,width=5cm,
  bbllx=280pt,bblly=70pt,bburx=510pt,bbury=770pt,angle=90.} 
\end{center}                                                                   
\caption[]{\label{newfig1}
\sl 
Smearing Acceptance, Stability and Purity at $Q^2$=20 GeV$^2$ for the 3 
kinematic reconstruction methods:
e (open triangles), DA (open squares), $\Sigma$ (closed circles).
}
\end{figure}

We distinguish two ways of defining the bin $\Delta$: one ($\Delta,t$)
based on the ``true'' kinematic variables, defined at the hadronic vertex;
the other  ($\Delta,r$) based on the reconstructed variables. We also 
consider the effect of the ``event selection cuts'' on the number of events 
($N_{\Delta }^c$) compared to the number of events before the cuts 
($N_{\Delta}$). These ``cuts'' are imposed to  improve the precision
of the measurement, but can  have an influence on the distribution
of the events in the kinematic plane, so the two cases must clearly be
separated. With these definitions, we  deduce from eq. (3) that
  $T_{\Delta}$ can be expressed as~\cite{pineiro} :
\begin{equation}
 T_{\Delta}= \frac {N_{\Delta,r}^c}{N_{\Delta,t}}=\epsilon_{\Delta,t} \cdot
A_{\Delta} \hspace*{1.cm}  \mbox{with} \hspace*{1.cm} 
\epsilon_{\Delta,t} \equiv \frac {N_{\Delta,t}^c} {N_{\Delta,t}} 
\hspace*{0.5cm}  \mbox{and} \hspace*{0.5cm} 
 A_{\Delta} \equiv \frac{N_{\Delta,r}^c}  {N_{\Delta,t}^c}
\end{equation}
The first term ($\epsilon_{\Delta,t}$) characterizes
 the  ``efficiency'' of the cuts in the bin $\Delta$, while the second one
can be defined as the ``smearing acceptance'' of the bin $\Delta$, 
since only the smearing of
the kinematic reconstruction is involved in its variations.
Since the cuts are choosen to have a high efficiency, we will not discuss
here the difference at the percent level which might occur between
$\epsilon_{\Delta}$ as determined on the data and on the simulation,
but focus on the behaviour of the smearing acceptance, which can have large
variations (up to hundreds of percents) across the kinematic plane, hence
determining the  measureable regions.
%The smearing acceptance variations are better studied rewriting $A_{\Delta}$ as
The smearing acceptance variations are better studied 
considering the ($\Delta,i$) subset which contains
the events belonging to  ($\Delta,i$) {\it and} ($\Delta,r$). 
Then  $A_{\Delta}$ can be rewritten as
\begin{equation}
A_{\Delta}= \frac {S_{\Delta}} {P_{\Delta}} \  \hspace*{0.7cm} \mbox{with} 
 \hspace*{0.4cm} S_{\Delta} \equiv \frac {N_{\Delta,i}^c} {N_{\Delta,t}^c} 
 \hspace*{0.7cm}   \mbox{and}  \hspace*{0.4cm} P_{\Delta} \equiv \frac {N_{\Delta,i}^c} 
{N_{\Delta,r}^c}
  \hspace*{0.5cm}  \mbox{;} \hspace*{0.5cm}
 N_{\Delta,i} \equiv N_{\Delta,t \cap \Delta,r} 
\end{equation}
The $S_{\Delta}$ and $P_{\Delta}$ are  refered as the stability and the
purity of the bin $\Delta$, since they characterize respectively:
\begin{itemize}
\item the proportion of genuine events of a bin, which are reconstructed in the
same bin. $S_{\Delta}$  characterizes  the number of events which migrate 
{\bf outside} of the bin $\Delta$.
\item the ratio  of genuine events of a bin  reconstructed in the
same bin divided by the total number of events reconstructed in that bin.
This last number is influenced by the number of events originating from
other bins  which migrate 
{\bf into}  $\Delta$.
\end{itemize}
The obvious goal is to maintain a stability and a purity as close
to unity as possible, hence $A_{\Delta}$ will also be close to 1.
However $A_{\Delta}$ can be close to unity even for  a low stability bin,
if its purity roughly matches its stability. In order to ensure a reliable
measurement of the cross-section, we will thus enforce conditions
on  $S_{\Delta}$ and  $P_{\Delta}$  separately, rather that only on their
ratio. To make a concrete example let us examine
the three standard reconstruction methods used
for DIS neutral currents at HERA, in the ($x,Q^2$) binning used by the
H1 collaboration, i.e 8 (5) bins per order of magnitude in $Q^2$ 
($x$)~\cite{H1F294}.
The smearing acceptance, stability and purity for the H1 detector
are shown in 
fig.~\ref{newfig1} for 
the electron, Double-Angle~\cite{stan} and $\Sigma$~\cite{bb}  methods,
in $x$ bins, at $Q^2$ = 20 GeV$^2$.
As mentioned, the acceptance can be equal   to unity, while the stability
and the purity are low ($\sim$ 25\%), like for instance at the lowest 
$x$ point in the DA method. Conversely, by requiring a ``reasonably'' high
stability and purity (we will use throughout this paper a minimum
of 30\%, as the H1
collaboration did for this binning~\cite{H1F294}), we ensure an 
acceptance close to unity. 

We observe in fig.~\ref{newfig1} that the stability has a more
regular behaviour than the purity or the acceptance, since it reflects only
the combination of the resolutions in $x$ and $Q^2$ in a given bin.
The stability can  be increased by enlarging the size of the bin.
Although the stability and purity are related quantities (they share the same
numerator), the purity is 
more irregular because it is influenced by events coming
from other bins, which may be populated in a different way, and have
different resolutions in the kinematic variables. Like the stability,
the purity can also be increased
by enlarging the bin size, however the migrations inside the bin depend
on several factors (resolutions in different bins, population of different
bins, structure function values in regions which are not ``measurable'',
radiative effects), 
rendering the control of the purity more delicate than the stability.
 Thus, in the following,
when comparing different methods, we will make the comparison on  their purity, but
only in the bins having a  minimum stability (chosen to be 30\%).

\section{Kinematic Reconstruction Methods at HERA}     
In the naive quark-parton model (QPM), the lepton scatters elastically
with a quark of the proton, and the two body final state is  
completely constrained using two variables, if we know the initial      
energies labeled E$_{\circ}$ and P$_{\circ}$ of the electron and proton.   
Similarly, the DIS variables can be determined using 2 independent variables,
which can be the energy (E) of the scattered electron, its polar 
angle\footnote{The positive $z$  
axis is defined at HERA as the incident proton beam direction.}($\theta$)
or  independent quantities reconstructed out of the hadronic final state
particles. For instance $\Sigma$, obtained      
as the sum of the scalar quantities $E_h-p_{z,h}$ 
of each particle (assumed to be massless) 
belonging  to the hadronic final state,  $p_{T,h}$
as its total transverse momentum or  the inclusive  
angle $\gamma$
 of the hadronic system\footnote{
We can define the similar quantities for the scattered electron: 
\\ \hspace*{1.1cm} 
$\sss=E~(1-\cos{\theta})$ 
%  \Sigma = \sum_{h} (E_h-p_{z,h})    
\hspace*{1.1cm}   $p_{T,e}=E\sin{\theta}$
\hspace*{0.6cm} i.e.  $\tan\frac{\theta}{2} = \frac{\Sigma_e}{ p_{T,e}}$.  } 
 which corresponds to the angle of the scattered quark
in the QPM:
%of the hadronic final state, and its total $\delta$ denoted $p_{T,h}$ and    
%$\Sigma$ in the following. 
%       
\begin{equation}   
  \Sigma = \sum_{h} (E_h-p_{z,h})    
  \hspace*{1cm}
  p_{T,h}   = \sqrt{(\sum_{h} p_{x,h})^2+(\sum_{h} p_{y,h})^2 }    
  \hspace*{1cm}
  \tan\frac{\gamma}{2} = \frac{\Sigma}{ p_{T,h}}      
\end{equation}     
$E_h,p_{x,h},p_{y,h},p_{z,h}$ are the four-momentum vector components   
of  each hadronic final state particle. 
$\Sigma$ is by construction minimally affected by  the 
losses in the forward direction
%\footnote{The positive $z$  
%axis is defined at HERA as the incident proton beam direction.}
due to the beam pipe hole in which  
the target jet and the initial state gluon radiation tend to   
disappear. $p_{T,h}$ covers the other spatial dimensions,    
and  is more sensitive to forward losses.
Thus to improve the kinematic reconstruction  we should  avoid using   
$p_{T,h}$ directly. 
We can use $\gamma$ instead, which carries the $p_{T,h}$ information      
and is better measured since in the ratio $\Sigma/p_{T,h}$  
the energy uncertainties cancel to first order   
and the effect of the losses in the forward    
beam pipe is diminished. Thus the optimal four ``detector oriented'' variables 
to characterize deep inelastic scattering at HERA are  
[E,$\theta,\Sigma,\gamma]$. 

Using these four input variables, there are 
three basic methods which make use of only two of them at a time, and which
are precise enough to allow sensible kinematic reconstruction,
namely: 

\noindent
- The electron only method ($e$) which uses E and $\theta$. \\
- The hadrons only method ($h$) which uses $\Sigma$ and $\gamma$~\cite{jb}\\
- The double angle method (DA) which uses $\theta$ and 
  $\gamma$~\cite{stan}.

In the following, all methods 
(for instance the DA) using some information from the  
hadronic final state will be called ``hadronic'', although
strictly speaking there is only one inclusive hadronic method
(a complete set of formulae is given in the appendix).
The $h$ method will not be discussed in the following since it is not 
precise enough compared to the others, for neutral current DIS events.

More than two variables are needed to determine the kinematics if the incident 
energy is unknown, which is the case when the incident electron 
emits a photon before the hard collision. This photon 
is  often undetected since it is emitted colinearly with
the incident electron beam direction, and can 
thus escapes inside the beam pipe.
In this case three variables are needed to  reconstruct the kinematics.
For instance $y_h={\Sigma} / {2E_{\circ}}$ can be  replaced by 
$ y_{\Sigma}\equiv {\Sigma} / {(\Sigma+\Sigma_e)}$  
to take into account the missing energy due to the escaping photon.
A further important characteristic of  $y_{\SSS}$ 
is that at high $y$, $\Sigma$ becomes the dominant term in the denominator 
and thus experimental errors on the $\Sigma$ measurement tend to cancel
between numerator and denominator. 
In the $\Sigma$ method~\cite{bb},
$Q^2_{\SSS}$ is constructed, like $y_{\Sigma}$  to be independent of 
QED initial state radiation (ISR) and 
to be optimal in terms of
resolution; thus  $p_{T,e}$ is used instead of  $p_{T,h}$:  
\beq
Q^2_{\SSS} \equiv p_{T,e}^2 / (1-y_{\SSS}) \hspace*{1cm}
x_{\SSS} \equiv  Q^2_{\SSS}/s y_{\SSS}.
\eeq 
Insensitivity to ISR  on $x_{\SSS}$  is achieved simply by
replacing  $s$ by  2~P$_{\circ}~(\SSS+\SSS_e)$, thereby obtaining
 the I$\SSS$
method which  is based on (E,$\theta,\SSS$). The DA method
was also  rendered ISR independent by using E 
to reconstruct the initial electron
beam energy. The IDA method~\cite{stan} obtained in this way is thus based
on  (E,$\theta,\gamma$).
The complete formulae and the comparison of these two methods can be found in~\cite{bb}. 
They will not be 
considered  in the following, since the gain obtained 
by the complete independence to ISR is not sufficient to compensate 
the loss of precision induced by the reconstruction of the incident electron
energy.

\begin{figure}[htb]                                                           
\begin{center}                                
\epsfig{file=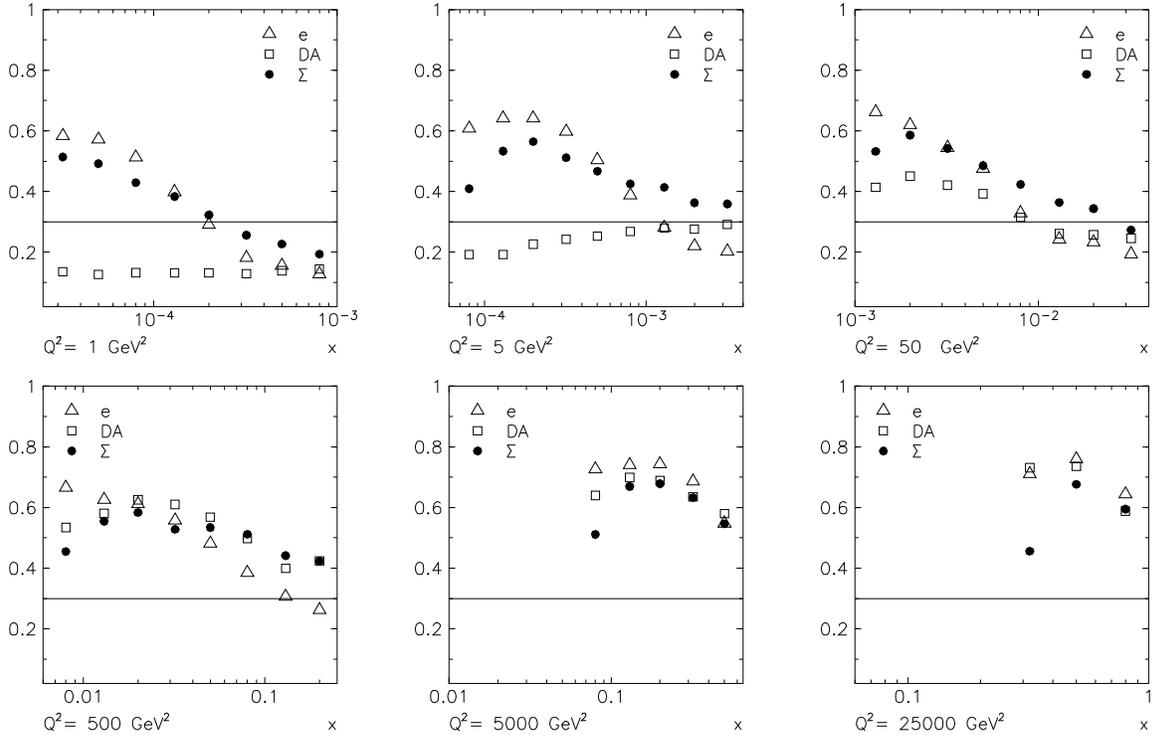,width=9.5cm,
  bbllx=70pt,bblly=70pt,bburx=510pt,bbury=770pt,angle=90.} 
\end{center}                                                                   
\caption[]{\label{newfig22}
\sl 
Purity at $Q^2$=1, 5, 50, 500, 5000 and 25000 GeV$^2$ for the 3 
kinematic reconstruction methods:
e (open triangles), DA (open squares), $\Sigma$ (closed circles).
}
\end{figure}

Fig.~\ref{newfig22} shows the purity  
of the $e$, DA and $\SSS$ methods as a function of $x$ in bins
of $Q^2$.  
The  properties of the 
3 standard methods to reconstruct the kinematics
of neutral current DIS events at HERA are clearly visible: 
high precision of the $e$ method at high $y$ with a severe 
degradation at low $y$, good precision for the
$\SSS$ method in the complete 
kinematic range,
high precision
of the DA method at medium ($\sim 10^2$ GeV$^2$) and high $Q^2$ 
({$\sim 10^4$ GeV$^2$)
 with a severe degradation at low $Q^2$ ({$\sim 1$ GeV$^2$).

The ``hadronic'' methods have already been shown to display
 at low $x$ and low $Q^2$ a rather imprecise
reconstruction  of $Q^2$. A simple solution to this problem
is to use $Q^2_e$  and to obtain,
via  $Q^2=xys$, either $x$  from a hadronic $y$
or $y$ from a hadronic $x$.
 For instance we have the  mixed method~\cite{max} 
($x_m$ is obtained from $y_h$ and $Q^2_e$) which has a good precision
at low $y$~\cite{gbwh,H1F293}. 
The precision at high $y$ can be improved by using the mixed $\Sigma$
($m\SSS$;  $y_{m\SSS} \equiv y_{\SSS}$ and $Q^2_{m\SSS} \equiv Q^2_e$), 
or better
the $e\SSS$ method ($x_{e\SSS} \equiv x_{\SSS}$ and $Q^2_{e\SSS} \equiv 
 Q^2_e$) as was shown on fig.~2 and 3 of ref.~\cite{bb}.

Another approach to combining two  methods which have complementary properties
has been tried by the H1 collaboration in the analysis of the diffractive 
structure function~\cite{H1DI94}. Since $Q^2=4 $E$_{\circ}^2(1-y) /
{\tan^2{\frac{\theta}{2}}} \ $
both in the $e$ and DA methods, 
an ``average'' method (labeled here ADA) has been
introduced, in which $y_{ADA}$ is 
obtained by weighting $y_e$ and $y_{DA}$ by $y$
and $(1-y)$ respectively:
\beq  
y_{ADA} \equiv y_e^2 + y_{DA}(1-y_{DA}) \hspace*{2cm}  
Q^2_{ADA} \equiv \frac{4 E_{\circ}^2(1-y_{ADA})} {\tan^2{\frac{\theta}{2}}}.
\eeq
On the non-diffractive DIS events
a better complementarity
is actually achieved in the A$\SSS$ method 
by ``averaging'' the $e$ and the $\SSS$ methods~\cite{andy}.
However neither the A$\SSS$ nor the  ADA method  bring 
an improvement compared to the
simpler $e\SSS$ method as can be 
seen in fig.~\ref{newfig3}: the A$\SSS$ method 
is giving similar performances
to the $e\SSS$ one except at low $x$ where it is slightly less precise, 
while the ADA method
is better at high $x$ and high $Q^2$, 
but weaker elsewhere, in particular at low $Q^2$.
\begin{figure}[htb]                                                           
\begin{center}                                
\epsfig{file=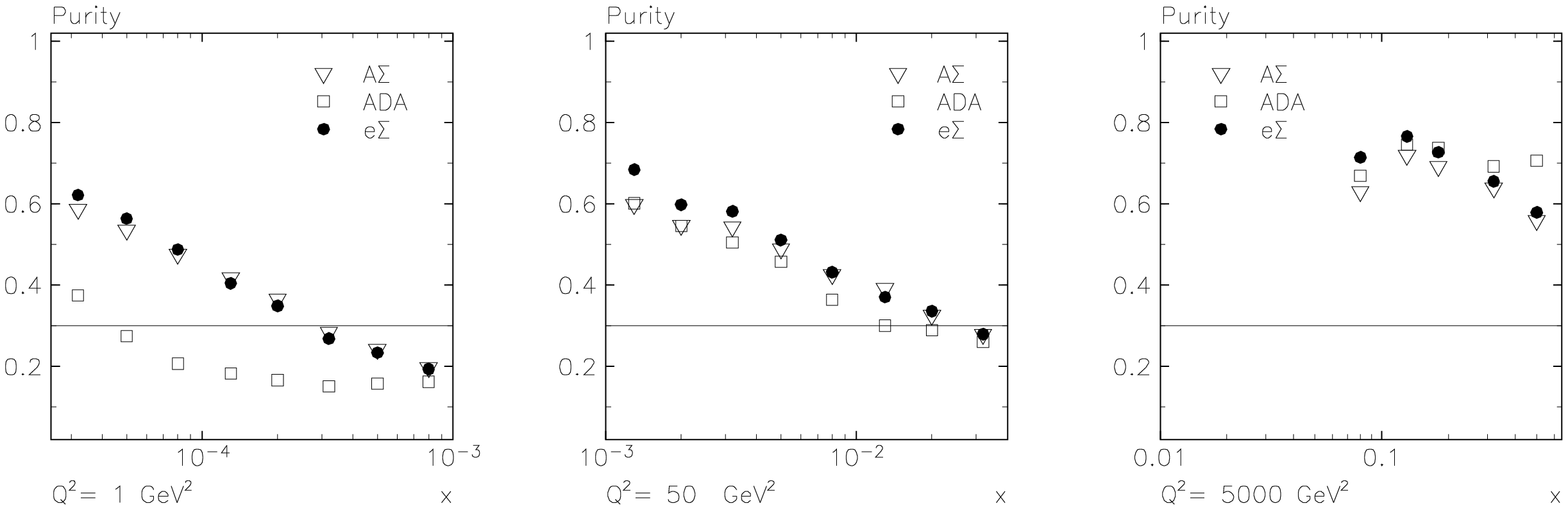,width=5cm,
  bbllx=280pt,bblly=70pt,bburx=510pt,bbury=770pt,angle=90.} 
\end{center}                                                                   
\caption[]{\label{newfig3}
\sl 
Purity at $Q^2$=1, 50 and 5000 GeV$^2$ for the 3 
kinematic reconstruction methods:
A$\SSS$ (open triangles), ADA (open squares), $e\Sigma$ (closed circles).
}
\end{figure}  

Before ordering these methods to underline their relations,
let us introduce another method derived from the  DA method, which will be 
useful later, for the understanding of more complicated methods and
for the study of error propagation. Following the
logic used for the $\SSS$ method,
the inclusive hadronic angle $\ggg$ can be replaced by $\gss$  
defined by $\tan {\gss} / {2}= {\SSS} / p_{T,e}$
since transverse momentum conservation implies $p_{T,h}=p_{T,e}$ even in 
the case of colinear ISR. 
This replacement improves the precision  on the hadronic angle
%since  $p_{T,e}$ is
%not affected, contrary to $p_{T,h}$, by particle losses in the beam pipe, and
%since $p_{T,e}$ is defined from the angle and energy of the electron which are
%experimentally reconstructed more precisely than their hadronic counterparts
but has the obvious drawback of introducing a  sensitivity to the 
electron energy reconstruction errors, which is absent in the DA method.
%, and thus less independent of the $e$ method).
This  D$\SSS$ method defined using $\gss$ instead of $\ggg$ in the DA 
formulae, gives:
\beq
y_{D\SSS} = \frac{\tan\gss/2}{\tan\gss/2
+\tan\theta/2}= \frac{\SSS/p_{T,e}}{\SSS/p_{T,e}+\sss/p_{T,e}}= y_{\SSS}.
\eeq
The $Q^2_{D\SSS}$ and $x_{D\SSS}$ are nevertheless different from their $\SSS$ 
counterparts 
(although  $x_{D\SSS}=x_m$) and we will see 
in the section 5 how the hadronic and electron  miscalibrations affect them.

In fig.~\ref{fig11}  the distribution of
the reconstructed $x$ ($x_{rec}$) divided by the true
$x$ ($x_{gen}$) and the equivalent distribution for $Q^2$
are compared for the $e$, DA, $\SSS$, D$\SSS$ methods at high $y$
(0.3-0.7)\footnote{
At low $y$, 
since all hadronic methods give rather similar results in terms of resolution,
(contrary to the $e$ method which strongly degrades),
the comparison is better made using the plots of the purity (fig.~2).
The purity of the D$\SSS$ method is almost identical to the $\SSS$ one  at
low $y$ (not shown).
}.
\begin{figure}[htb]                                                           
\begin{center}                                                                 
\epsfig{file=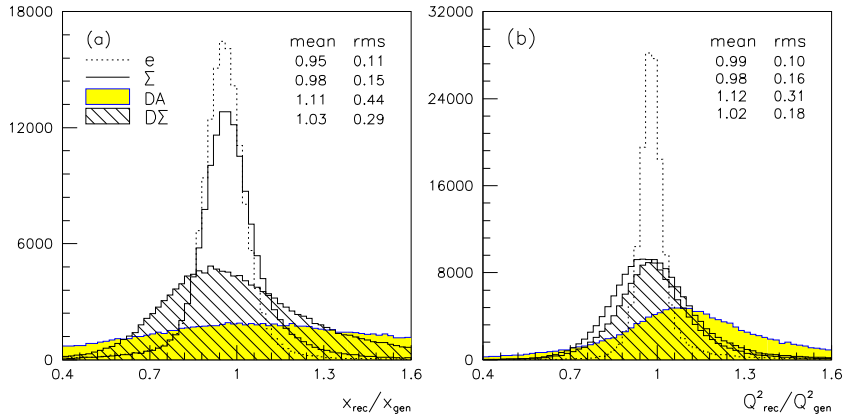,width=11cm,
  bbllx=50pt,bblly=440pt,bburx=530pt,bbury=650pt} 
\end{center}                                                                   
\caption[]{\label{fig11}
\sl Distribution of  $x_{rec}/x_{gen}$ (a) and $Q^2_{rec}/Q^2_{gen}$ (b)
for the $e$, DA, $\SSS$ and  D$\SSS$ methods  for $Q^2 > 7$ GeV$^2$ at
high $y$ ($0.3<y<0.7$)
}
\end{figure}        
This figure, as all the figures in this paper,  is obtained from a 
deep inelastic sample simulated in detail in the H1 detector (1995  set-up)
as described in~\cite{H1F295}. All known effects
both from the detector  and from the physics point of view 
(structure functions,
QED radiation)
are included  in the simulation, which has been shown to give a good 
description 
of the H1 data in the complete kinematic range~\cite{H1F295,H1F296,HIQ2F2}.
The uncertainty on the electron energy, on the polar angle and on the 
hadronic energy
scale are typically 1\%, 1~mrad and 4\% respectively. For figs. 4 and 5,
a cut at  $Q^2 \ge 7$ GeV$^2$ has been made, which corresponds at HERA to the
region measured with the highest precision. In fig.~4 the excellent resolution
at low $x$ and low $Q^2$ of the $e$ method is clearly visible. Only the
${\SSS}$ method  has a comparable, although lower,
 resolution in $x$. In $Q^2$ the 
weakness of all these  hadronic methods is visible. The D$\SSS$ method
is more precise than the DA one at low $Q^2$, 
as expected from the way it is constructed. Unfortunately the $e$ method
is not precise at low  $y$  (as can  be seen in fig.~2), so either
the structure function 
measurement is done using different methods in different 
regions of the kinematic plane, or a precise method over the complete
kinematic plane is found. Both these approaches have already been used
at HERA, but for the  more precise future
measurements it is mandatory to focus on the second case and to
optimize the reconstruction method to be used.

To conclude  this section,
the hadronic  methods are ordered in table~\ref{tab1}: 
each column represents the
type of approach followed (Sigma, Double-Angle, Mixed), 
while each row represents a prescription to derive them:
basic method, ISR independent, ``optimized resolution''.
The relations become clear and  the effects coming from the type of 
recontruction used can be disentangled, by comparing methods 
from  the same row, from those due to 
the detector response which can be studied  
by comparing methods from  the same column.
\begin{table}
\begin{small}
\centering
\begin{tabular}{|c|c|c|c|}
\hline 
   & Sigma & Double-Angle & Mixed \htab\\
\hline 
basic method &       $h$ & DA \ ($\ggg$) & $m$ \htab\\
\hline
ISR independent & I$\SSS$ & IDA \ ($\ggg$) & $m\SSS$ \htab\\
\hline
optimized method & $\SSS$ & D$\SSS$ \ ($\gss$) & $e\SSS$ \htab\\
\hline
\hline 
rescaled method & $r\SSS$ & PT \ ($\gamma_{r\SSS}$) & $re\SSS$ \htab\\
\hline
\end{tabular}
\caption  {\label{tab1}        
\sl Classification of the hadronic methods. 
For the methods using the Double-Angle approach, the correponding hadronic
angle is also given.
Note that the
$m\SSS$ method is ISR independent only in $y$. The methods displayed in the 
last row are discussed in section 4.
}
\end{small}
\end{table}

\section {Kinematic Reconstruction Improvements}
The  description of the methods given above 
suggests that it might be possible to further optimize
the reconstruction precision. The most direct way would be to perform a fit
 of the kinematics using all information available as described for instance 
in~\cite{chaves,julian}. However the redundancy is not large, since
$p_{T,h}$ is not precise enough to be helpful in neutral current events,
except at  high $Q^2$.
Furthermore the
method requires a good knowledge of the uncertainties of the measured
variables over all the kinematic plane and the lack of statistics has prevented
the production of a detailed enough map of
all imperfections of the detectors. 
%
%simulation have so far prevented the kinematic fitting
%approach being used at HERA.
We will thus concentrate on the ``analytic'' improvement of the hadronic 
methods, in particular by  ``rescaling''  the hadronic energy:
%as was done
%by the ZEUS collaboration
%in the so-called ``PT'' method used for the structure function analysis
%of the 1994  data~\cite{ZEUSF294}.
%
%What is meant by  ``rescaling''
%is that 
$\SSS$ may be rescaled to approach its true value
using the formula $\SSS+\sss=2$ E$_{\circ}$,  if the error on $\sss$
is assumed to be small compared to the $\SSS$ 
one~\footnote{Indeed, the relative error on $\SSS$ increases when going to  
 lower $x$ and lower $Q^2$  since 
the hadronic final state contains an increasing
fraction of low energy particles which are measured less
precisely and which may go undetected in the HERA
detectors.}.
%\footnote{The measurement of the hadronic final state using
%simultaneously the tracking devices for low momentum 
%charged particles and the calorimeter for the other particles,
%as done in the H1 collaboration~\cite{H1F294,bas} 
%brings a sizeable
%improvement by effectively reducing the energy threshold and allows 
%to detect the particles down to 100 MeV.}.
%This procedure has the drawback 
%to postulate that E$_{\circ}$ is the incident electron
%energy, which is not  true in case of ISR, but as in the comparison 
%of the $\SSS$ and 
%the I$\SSS$ method, the drawback is smaller than the gain in resolution
%once the radiative processes are simulated precisely.
But 
if $\Sigma$ is expressed as $2$ E$_{\circ}-\sss$ in the $h$ or $\SSS$ method,
these methods become identical 
to the $e$ method. An intermediate solution between
the $\SSS$ and $e$ method can be obtained by using
the rescaling factor defined as 
\beq 
r \equiv \frac{2 E_{\circ}} {\SSS+\sss} 
\eeq
as was already implicitly done  in some of the previously discussed
variables:
\beq
y_{\SSS}= r \ y_h \ ; \hspace*{1cm}  
y_{e\SSS}= r \ y_{\SSS} \ ; \hspace*{1cm}  
Q_{\SSS}^2= \frac{Q^2_e }{r}\ ; \hspace*{1cm}  
Q_{D\SSS}^2= Q^2_e \ r\ .
\eeq
An  improvement on the $\SSS$ kinematic variables at low $x$ is
obtained by directly rescaling $\SSS$, thereby
 defining an $r\SSS$ method:
\beq
y_{r\SSS} \equiv \frac{r \SSS} { r\SSS+\sss} \hspace*{2cm} 
Q_{r\SSS}^2 \equiv  \frac{p_{T,e}^2} {1-y_{r\SSS}}  
\eeq
The corresponding ``rescaled'' hadronic angle is defined as 
$\tan{\ggg_{r\SSS}} = r\SSS 
/ p_{T,e}$ and allows to define a rescaled D$\SSS$ method ($r$D$\SSS$).
This method was already derived in a different way  by the ZEUS collaboration,
and is called the ``PT'' method\footnote{
Actually, the  ``PT" method includes also a calibration of the hadronic
final state using  $p_{T,e}$. Here we single out the analytic definition
of the method.}~\cite{ZEUSF294}. Similarly an $re\SSS$ method is defined, using
$x_{r\SSS}$ and $Q^2_e$.
With these definitions we create the last row of table~\ref{tab1}
 with the  ``rescaled" methods.
%\begin{center}
%\begin{tabular}{|c|c|c|c|}
%\hline 
%rescaled method & $r\SSS$ & $r$D$\SSS \equiv PT$ 
%\ ($\gamma_{r\SSS}\equiv \gamma_{PT}$) & $re\SSS$ \htab\\
%\hline
%\end{tabular}
%$r\SSS$, $r$D$\SSS$ and $re\SSS$.
%\end{center}
These new methods are compared to the $e$ and $\SSS$ methods in $x$ and
$Q^2$ in fig.~\ref{fig12} in the same $Q^2,y$ intervals as those of 
fig.~\ref{fig11}.
\begin{figure}[htb]                                                           
\begin{center}                                                                 
\epsfig{file=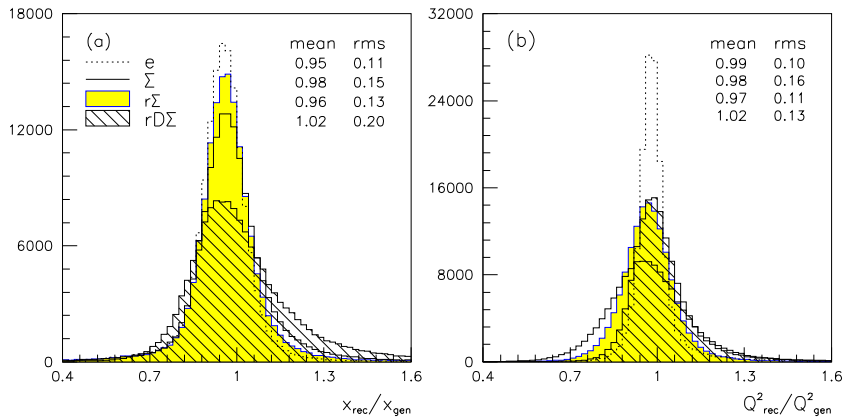,width=11cm,
  bbllx=50pt,bblly=440pt,bburx=530pt,bbury=650pt} 
\end{center}                                                                   
\caption[]{\label{fig12}
\sl Distribution of  $x_{method}/x_{gen}$ (a) and $Q^2_{method}/Q^2_{gen}$ (b)
for the $e$, $\SSS$, $r\SSS$  and  $r$D$\SSS$ ($\equiv$PT) methods 
for $Q^2 > 7$ GeV$^2$ at
high $y$ ($0.3<y<0.7$).
}
\end{figure} 
\noindent       
With the rescaling, $x_{PT}$ ($x_{r\SSS}$) has indeed a better resolution than
$x_{D\SSS}$ ($x_{\SSS}$). However $x_{PT}$ is still less precise 
for $y$ values between 0.3 and 0.7. than
the simple, non-rescaled, $x_{\SSS}$, due to the propagation of the hadronic
error in these two methods (cf section 5). 
The difference in $x$ between the  rescaled methods and the $e$ method
is now smaller, which is also true in  $Q^2$ where
the  rescaled methods ($r\SSS$ and PT)
are similar and definitely better than the non-rescaled ones
($\SSS$ and D$\SSS$). 
%We can conclude from this short study of rescaling, that if considering
%only hadronic uncertainties, the $r\SSS$ method has better resolution
%in $x$ and same resolution in $y$ or $Q^2$ than the PT ($r$D$\SSS$)
%method. 
By using the $Q^2_e$, the $re\SSS$ combination 
allows  a further improvement  
with respect to the two other rescaled methods or to the $e\SSS$ method.
The overall behaviour in the complete kinematic plane of one of the rescaled
methods (the PT), compared to the $e\SSS$ can be judged by their purity
shown  in fig.~\ref{newfig66}. 
They both show a precise behaviour over the complete kinematic plane.
The $e\SSS$ method  is more precise at low $x$, at high $Q^2$,
and has a comparable or better purity 
than the $e$ method (compare fig.~\ref{newfig22} and fig.~\ref{newfig66})
even at low $x$ and low $Q^2$. 
\begin{figure}[htb]                                                           
\begin{center}                                
\epsfig{file=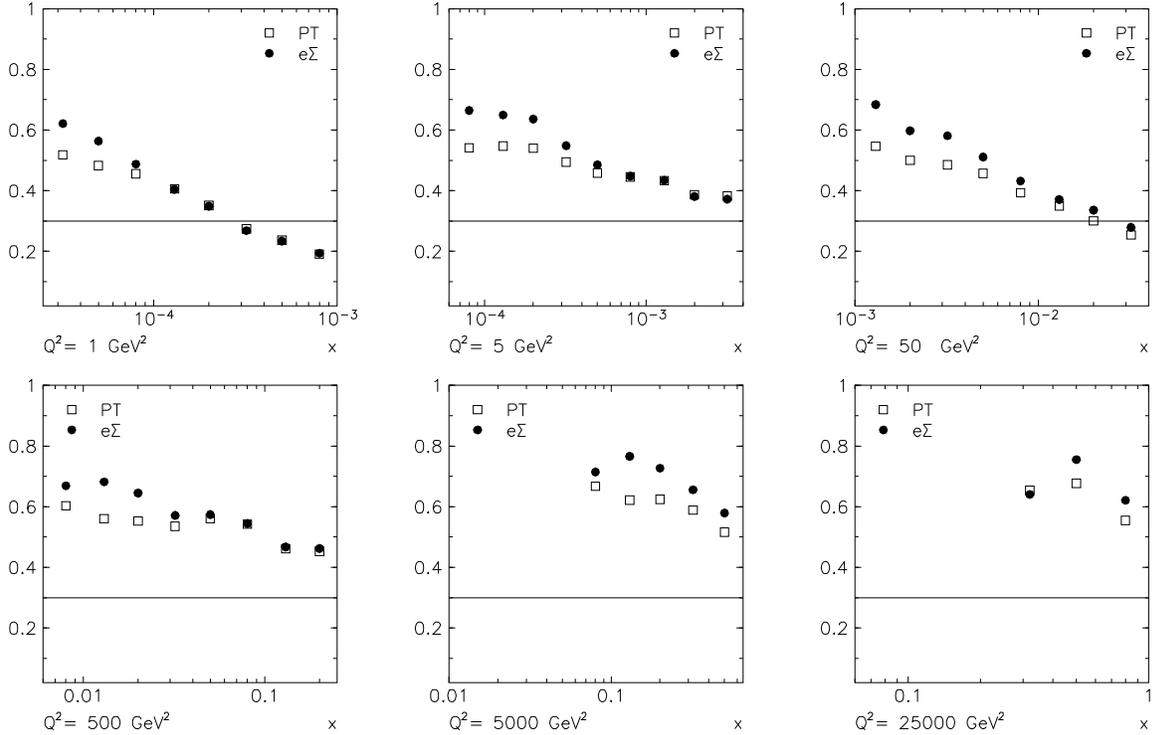,width=9.5cm,
  bbllx=70pt,bblly=70pt,bburx=510pt,bbury=770pt,angle=90.} 
\end{center}                                                                   
\caption[]{\label{newfig66}
\sl 
Purity at $Q^2$=1, 5, 50, 500, 5000 and 25000 GeV$^2$ for the PT (open squares)
and $e\SSS$   (closed circles) 
kinematic reconstruction methods. With the high precision measurement
of the  electron energy in the  H1 detector,
the $e\SSS$ method is better than the PT method
at low $x$ and also at high $Q^2$ (see fig.~7 for further comments on this
matter).
}
\end{figure}        
%
%The resolution in $y$ at high and low $y$ are compared in fig.3 between 
%$y_e, y_{\SSS}, y_{r\SSS}=y_{p_T}$ and  $y_{re\SSS}$. 
%The $y_{e\SSS}$ resolution
%(not shown) is very similar to the $y_{r\SSS}$ one.
%
%As already mentioned 
%the rescaling by $r$ introduces  a dependence on ISR, which was not present
%in $y_\SSS$ and $Q^2_{\SSS}$. The propagation of this effect is more
%pronounced 
%in the $r$D$\SSS$ method than in the $r\SSS$ one, since the ISR introduces
%shifts of the same sign in $Q^2_{r\SSS}$ and $y_{r\SSS}$
%which tend to cancel in $x_{r\SSS}$, and to add up in $x_{rD\SSS}$ .

Before examining in the next section the propagation of the errors 
on the kinematic variables,
let us consider the other characteristic of the rescaling, i.e.
the increased dependence on the electron variables,
since the rescaling factor  depends
also on the reconstructed energy and angle 
of the scattered electron.      
%The effect of the miscalibrations of these two
%variables are studied in the next section but here 
We can indirectly 
demonstrate the increased dependence by applying a further rescaling, and 
obtain three new methods: $r_2\SSS$, $r_2$D$\SSS$ and $r_2e\SSS$.
The $r_2$ rescaling factor is obtained recursively 
(assuming $r_0=1, r_1=r$ and  $y_{r_0\SSS}=y_{\SSS}$) using the 
following formulae:
%\beq
%y_{\SSS}= \frac{y_{h}} {y_{h}+1-y_e} \hspace*{1cm} 
%y_{r\SSS}= \frac{y_{\SSS}} {y_{\SSS}+1-y_e} \hspace*{1cm} \mbox{and} \hspace*{1cm} 
% r = \frac{2 \ E_{\circ}} { r_{0}\SSS+\sss}
%\eeq
%We can thus define  $y_{r_n\SSS}$ and  the $r_{n}$ rescaling
% factor by the relation  
\beq
y_{r_n\SSS}= \frac{y_{r_{n-1}\SSS}} {y_{r_{n-1}\SSS}+1-y_e}
\hspace*{1cm} \mbox{and} \hspace*{1cm} r_n= \frac{2 \ E_{\circ}} { r_{n-1}\SSS+\sss}
\eeq
The resolution at high $y$ improves at each rescaling and
actually
$y_{r_n\SSS} \rightarrow y_e$ when $n \rightarrow \infty$\footnote{Since 
$y_{r_n\SSS}$ can be written as $\frac{1}{1+u+u^2+..+u^{n-1}+u^n/y_{\SSS}}$,
with $u \equiv 1-y_e$} implying that
the three approaches ($r_n\SSS, r_n$D$\SSS$ and $r_ne\SSS$)
converge to the $e$ method.
This means that the gain in precision at high $y$ is obtained at the  
expense of a loss of precision at low $y$.
Furthermore the ``hadronic" method becomes more influenced by the $e$ errors
and the significance of 
the cross-check of the systematics which can be made when using the
$e$ and $\SSS$ methods independently is reduced at each rescaling. 
However, we will now study the influence 
of the hadronic and electron reconstruction errors 
on these methods, to see when the usage of a
rescaled method  is worthwile.

\section{Error Propagation on the Kinematic Variables}      

In this section we will first
consider that the hadronic final state reconstruction
is the dominant source of systematic error on the kinematic variables,
and thus neglect the effect of the electron reconstruction, which will be
considered specifically in the second part of the section.
This approximation is  legitimate at HERA, in particular at low
$Q^2$, since the energy carried by the hadronic final state which is
visible in the detector becomes on average smaller as $Q^2$ decreases.
Since the hadronic final state is a combination of several particles
which enter the detector in different places, the precise effect of the 
detector response can be obtained only via a complete detector 
simulation/reconstruction program. 
However we can start by examining  the propagation
of the observable errors on the kinematic variables,
before comparing the  results to a realistic simulation.
Namely we will consider errors 
like $\delta \SSS / \SSS$ and $\delta p_{T,h} / p_{T,h}$
which are related to the error on the hadronic angle  $\delta \ggg$ by
\footnote{In all the error equations, negative signs may appear in front
of some terms, allowing to know the direction of the bias introduced
by a given measurement error. When the terms are independent, 
the actual total error squared  is obtained 
as usual, i.e. by quadratically summing the different terms.}
\beq
 \frac
{\delta \ggg }{\sin \ggg} = \frac{\delta \SSS} { \SSS} - \frac{\delta p_{T,h}} { p_{T,h}}
\eeq
%If the errors on $\SSS$ and $p_{T,h}$ would come only from a constant energy
%miscalibration then the error on $\ggg$ would be zero. 
The error on $\gamma$ arises from 
the spread of the hadronic final state particles in the detector, 
since it is not possible to ensure complete
homogeneity in the detector response (even with a perfect absolute calibration,
and with a perfect angular reconstruction, the fluctuations in 
the energy measurement have a different influence 
on $\SSS$ and $p_{T,h}$ as soon as there is
more than one particle in the final state). 
In addition particle losses in the beam pipe affect $\SSS$ and
$p_{T,h}$ in a different way: losses in the forward (backward)
beam pipe affect to a good approximation mainly $p_{T,h}$ ($\SSS$).
This explains why, at low $Q^2$,
 the DA method is not precise both at very low  and
very high $y$.  The complete error propagation of $\delta \SSS$ and
$\delta \ggg$ for $x, y$ and $Q^2$ for the $h$, DA and $m$  methods 
are given in the appendix, but are not discussed further here since
these methods 
 do not allow a precise measurement over the full kinematic range.
%At high $y$ the DA method allows for a better precision than the $h$ method
%as a $(1-y)$ term multiplies $\delta y/y$. 
%The resolution in $x$ is nevertheless
%better in the $m$ method, since the error on $Q^2_e$ is small  compared to
%the one of $Q^2_{DA}$.

For the optimized and for the rescaled methods, 
the corresponding error propagation 
are given in table~\ref{tab2},
noting that they depend only on one hadronic variable, 
$\Sigma$. 
\begin{table}
\begin{small}
\centering
\begin{tabular}{|c||c|c|c|c|c|}
\hline 
    & $\SSS$ &  D$\SSS$  &    $e\SSS$   & 
PT ($\equiv r$D$\SSS$)    & $r\SSS$ \htab\\ 
\hline 
\hline 
      \dyy &
      $(1-y)$  \dss &
      $(1-y)$  \dss &
      $(1-2y)$ \dss &
      $(1-y)(1-y_{r})$   \dss &
      $(1-y)(1-y_{r})$ \dss 
                            \htab\\
\hline
      \dqq &
      $ y$  \dss  &
      $-y$  \dss  &
      $ 0 $ & 
      $-y_{r} (1-y)$  \dss &
      $ y_{r} (1-y)$  \dss 
                           \htab\\
\hline
      \dxx &
      $(2 y-1)$  \dss &
      $ -y$      \dss &
      $(2 y-1)$  \dss &
      $     -$\dss &
      $(2 y_{r}-1)(1-y)$ \dss  
                              \htab\\
\hline
\end{tabular}
\caption  {\label{tab2}        
\sl Errors on $y, Q^2$ and $x$ due to errors on the hadronic observables 
for the $\SSS$, D$\SSS$, $e\SSS$, PT (=$r$D$\SSS$) and  $r\SSS$ methods. 
For the coefficients in this table, the following abreviated notations
are used: \hspace*{0.4cm}
$y \equiv y_{\SSS}$ ; \hspace*{0.4cm} $y_r \equiv y_{r\SSS}$ .
}
\end{small}
\end{table}
At low $y$ ($y \le 0.1$), the errors of 
the three methods become identical:
\beq 
   \frac{\delta y}{y} \simeq \frac{\delta \Sigma}{\Sigma} \hspace*{1cm}
   \frac{\delta Q^2} {Q^2} \simeq 0 \hspace*{1cm}
   \frac{\delta x} {x}  \simeq -\frac{\delta \Sigma}{\Sigma}
\eeq
and a more precise study of the effect of the 
losses in the  forward beam pipe and of
the $p_{T,e}$ reconstruction uncertainties
is needed  since they become more important than those due
to the hadronic energy uncertainty.

At high $y$ the error originating from the hadrons dominates and
the situation is drastically changed: 

\noindent
$\bullet$ 
For the optimized reconstruction methods, the $\SSS$ approach allows
smaller errors than the DA or  D$\SSS$ approach, as can be seen by comparing 
$\delta x_{\SSS} /x_{\SSS} $ to $\delta x_{D\SSS} /x_{D\SSS}$ in the
$y$ interval $0.3-0.7$. The computed average 
ratio of these errors is 1/3 and is even smaller
 if comparing the $\SSS$ to the DA method. This explains the
results already discussed of  fig.~\ref{fig11}.

\noindent
$\bullet$ 
For the rescaled methods (PT and $r\SSS$)
the errors on $x$ and $Q^2$, with respect to the corresponding non-rescaled 
method are reduced by a factor $(1-y)$, providing an improvement
at high $y$. 
Both methods have a reconstructed $Q^2$ which
deviates from the true $Q^2$ by the same amount but
in opposite directions. 
For $x$ we have a similar situation as for the 
non-rescaled case since
the improvement by a factor of $(1-y)$ in both methods does not 
cancel the difference in resolution observed above  between 
$x_{\SSS}$ and $x_{D\SSS}$, so the ${r\SSS}$ (and actually also the
simple $\SSS$) method provides
a  better estimation of $x$ than the PT one.
So, when the hadronic error is the dominant one, a method of the Sigma
type should be chosen  to have the best kinematic reconstruction 
over the widest kinematic range.

As already said, 
the ``hadronic" methods described above are also influenced by the
reconstruction of the electron angle and energy. 
Here we examine the situation at low $Q^2$ in which typical values
for  $\delta E/E$  and $\delta \theta$ in the  HERA experiments are 
1\% and 1 mrad. The case of high $Q^2$ events is treated in section 6. 
From  the error propagation of E and $\theta$ on the
kinematic variables reconstructed with the $e$ method, we can observe that
the $\theta$ error becomes significant only when $\theta$ tends to its
extreme (0 and 180$^{\circ}$). At low $Q^2$, i.e. if 
$\theta$ is greater than about $175^{\circ}$
its error  dominates the total error.
For the following discussion  we do not consider  $\delta \theta$
if it is not weighted by a $\tan \theta/2$ factor large enough
to render this error comparable to the error originating 
from the electron  energy
miscalibration.  Considering only the uncertainty due to the angular
measurement, we have:
\beq
\dye \simeq 0 \hspace*{2cm} \dqe \simeq -\tan \frac{\theta}{2} \ \delta\theta 
\hspace*{2cm} \dxe \simeq -\tan \frac{\theta}{2} \ \delta\theta
\eeq
For the DA method we have a different $\theta$ dependence (see table 5):
\beq
{\dayy} = {-\frac{1-y_{DA}}{\sin{\theta}}} \  \delta \theta     \hspace*{1cm}
{\daqq} \simeq  {-\frac{2}{\sin{\theta}}} \  \delta \theta     \hspace*{1cm}
{\daxx} \simeq {-\frac{1+y_{DA}} {\sin{\theta}}} \  \delta \theta    
\eeq
implying a more severe degradation at low $Q^2$ than for the $e$ method,
since $y_{DA}$ is strongly
affected by the $\theta$ error, and $x_{DA},Q^2_{DA}$ are more affected
than  $x_{e},Q^2_{e}$.

To study the other ``hadronic" methods
we   observe that the errors on $\sss$ and $r$ due
to $\delta \theta$ are such that:
\beq
\delta \Sigma_e / \Sigma_e \simeq 0  \hspace*{1cm} \mbox{;} \hspace*{1cm}
\delta r / r \simeq 0 
\eeq
implying
that for all hadronic methods which use $p_{T,e}$ ($\SSS, $D$\SSS, r\SSS$,
PT, $e\SSS$,..)
the errors due to $\theta$ (see also table 6) are identical, namely:
\beq
{\ddyy} \simeq 0 \hspace*{2cm} {\ddqq} \simeq -\tan \frac{\theta}{2} \ \delta\theta 
\hspace*{2cm} {\ddxx} \simeq -\tan \frac{\theta}{2} \ \delta\theta
\eeq
We thus have a simple situation: the error on $\theta$ propagates in
the same way in the $e$ and in all these hadronic methods and this fact
comes from the usage of the transverse momentum of the electron.
Since $y$ is essentially not affected by the $\delta \theta$ error,
similarly all hadronic $y$ are also independent of this error.
However, both $Q^2$ and $x$ have a divergent error at large angle, implying 
that the measurement of the lowest 
$Q^2$ by using larger  angles is hampered
by the angle error propagation. This suggests the usage of lower
incident (beam) electron energies to measure precisely the  lower $Q^2$ domain,
rather than further increasing  the acceptance at large angle as has
already been achieved at HERA, by shifting the interaction
position by about 70 cm in the forward direction and by upgrading the
detectors close to the backward beam pipe.

To study now the influence of the electron energy E on the standard and
rescaled methods,
we compare in table~\ref{tab3} the energy error
\begin{table}[htb]
\begin{small}
\centering
\begin{tabular}{|c||c|c|c|c|c|}
\hline 
    & $\SSS$ &  D$\SSS$  &    
$e\SSS$   & 
%$e$   & 
PT ($\equiv r$D$\SSS$)    & $r\SSS$ \htab\\ 
\hline
\hline
      \dyy &
      $ (y-1)$    \dee &
      $ (y-1)$    \dee &
      $ 2(y-1)$  \dee &
%%      $ \frac {y-1}{y}$  \dee &
      $  (y_{r}-1) (2-y)$ \dee &
      $  (y_{r}-1) (2-y) $  \dee 
\htab\\
\hline
      \dqq &
      $ (2-y)$ \dee &
      $    y$  \dee &
                        \dee &
      $  y_{r} (2-y)  $  \dee &
      $  2-y_{r} (2-y)$ \dee
\htab\\
\hline
      \dxx &
      $(3-2y)$           \dee &  
                         \dee &  
      $(3-2y)$           \dee &    
%%      $\frac{1}{y_e} $           \dee &
      $(2-y) $           \dee &
      $ 2- (2 y_{r}-1)(2-y)$    \dee 
\htab\\
\hline
\end{tabular}
\caption  {\label{tab3}        
\sl Errors on $y, Q^2$ and $x$ due to errors on the energy of the electron 
for the  $\SSS$, $D\SSS$, $e\SSS$, PT  and  $r\SSS$  methods.
The following notations
are used: \hspace*{0.2cm}
$y \equiv y_{\SSS}$ ; \hspace*{0.2cm} $y_r \equiv y_{r\SSS}$.
 }
\end{small}
\end{table}
propagation  for the  $\SSS$, D$\SSS$, $e\SSS$, PT and $r\SSS$ methods,
keeping in mind for comparison
the well-known $\frac{1}{y}$ \dee $ \ $ behaviour of $\delta x_e/ x_e$.  
The lack of dependence 
on E in the DA method is  changed  to a weak dependence
in the D$\SSS$ method. The E dependence of $Q^2$ 
for the $\SSS$ method (and to a lesser extent for the $e$ method: $\delta 
Q^2_e/Q^2_e = $ \dee),
is stronger than the D$\SSS$ one. At  high $y$ however,
the influence of E on the $Q^2$ of these three methods become similar.
A similar effect is visible on $x$, for which the E dependence
of the D$\SSS$ is small but increases at high $y$, while it is larger
for the $\SSS$ and $e$ methods at low $y$  but decreases with increasing 
$y$.
For the rescaled methods the same message holds, but the E dependence 
on the PT method is larger than on the D$\SSS$ one (a factor 
$2-y$ larger, for all variables), and conversely,
smaller for the   $r\SSS$ than for the $\SSS$ one (from $\propto 2-y$ to
$\propto 2-2y$), thereby reducing the initial difference between the
$\SSS$ and D$\SSS$ methods.

As a general conclusion, the PT method 
is better suited for low $y$ than for
high $y$ and is less influenced by the electron energy miscalibration.
The usage of the PT method should thus be reserved to the case
in which the miscalibration error on the electron is of the same
order as the hadronic one. This can be verified by comparing
fig.~\ref{newfig7}a,
which shows the purity at $Q^2$=20 GeV$^2$ of the $e\SSS$ and
of the three rescaled methods in the nominal calibration conditions
of the H1 experiment, to fig.~\ref{newfig7}b, which 
\begin{figure}[htb]  
\begin{center}                                
\epsfig{file=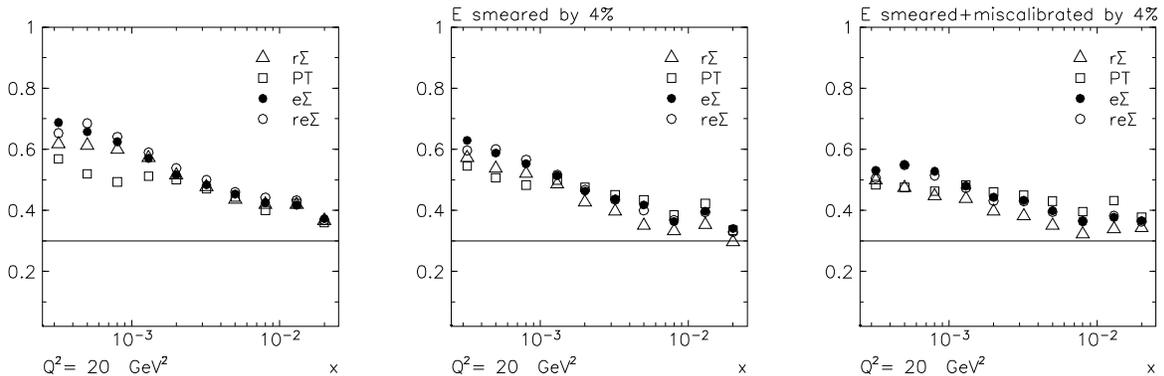,width=5cm,
  bbllx=280pt,bblly=70pt,bburx=510pt,bbury=770pt,angle=90.} 
\end{center}                                                                   
\caption[]{\label{newfig7}
\sl 
Purity  at $Q^2$= 20 GeV$^2$ for  4 
kinematic reconstruction methods:
$r\SSS$ (open triangles), PT (open squares), $e\Sigma$ (closed circles) and
 $re\SSS$ (open circles). a) nominal calibration; b) Electron energy
smeared by 4\%; c) Electron energy
smeared and miscalibrated by 4\%.
}
\end{figure}  

\noindent
shows the same purities,
but obtained after degradation of the electron energy response with a gaussian
smearing of 4\%, and to fig.~\ref{newfig7}c, which has an additional systematic
shift on the electron energy scale of $-4\%$. 
The figures show how the PT method remains relatively insensitive to the
energy degradation, thereby becoming more precise than the other methods
which are more affected by the degradation.
%This is not true  for the $\SSS$ (or $r\SSS$) which is precise
%also at high $y$ but for a  relatively poor $Q^2$ resolution
%Using $x_{\SSS} (or x_{r\SSS})$  and  {\bf $Q^2_e$} instead of 
%$Q^2_{\SSS}(or Q^2_{r\SSS})$ allows this problem to be cured and
%to have high precision  over the
%full $x$, $y$  and $Q^2$ range.

\section{Kinematics at High $Q^2$}
With the foreseen increase of 
luminosity, the HERA physics program in the next decade will
focus on physics at large momentum transfer. These studies have been already
started, and have been spurred by the observation of an excess of
NC DIS events at high $Q^2$ 
($Q^2 > 15000$ GeV$^2$)~\cite{h1hiq296,zeushiq296}. 
The difficulty of
reconciling both the H1 and ZEUS signals \cite{bb2,drees} 
in the hypothesis of  a single narrow resonance
has underlined the importance of kinematic reconstruction at high $Q^2$
in particular when dealing with low statistics on which radiative
effects can have a strong influence. For the future data, 
a simple comparison of the different methods can be made, based on
the bin purities  and using high simulated
statistics. A more elaborate study would require for instance to study
the optimal ($x,Q^2$) binning for each method. However the basic
properties can be understood with a single binning choice well 
adapted to the goal pursued by the two collaborations, i.e. an uncertainty
on the electron and hadronic energy measurement 
of about 1 and 2 \%  respectively. 

We compare in selected high $Q^2$ bins, in fig.~\ref{newfig8}, the
optimal $e\SSS$ method 
to the two simple  methods which are precise
at high $Q^2$, i.e. the $e$ and DA ones. 
The DA method has a good purity
in the whole ($x,Q^2$) range displayed, but is in general less precise
than the $e$ or $e\SSS$ method. These two last methods display a  similar
behaviour at high $Q^2$, both at high and low $x$. At $Q^2$ below 
10000 GeV$^2$, the low $y$ weakness of the $e$ method starts to be 
visible in the highest $x$ points (see also fig.~2). 
\begin{figure}[bth]  
\begin{center}                                
\epsfig{file=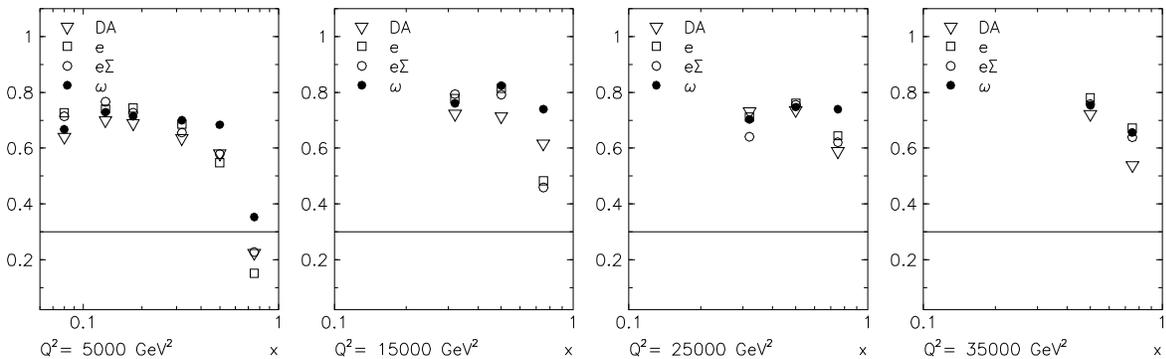,width=5cm,
  bbllx=280pt,bblly=70pt,bburx=510pt,bbury=770pt,angle=90.} 
\end{center}                 
\caption[]{\label{newfig8}
\sl 
Purity at high $Q^2$  for  4 
kinematic reconstruction methods:
$e$ (open squares), DA (open triangles), $e\Sigma$ (open circles) and
 $\omega$ (closed circles).
}
\end{figure}  

These three methods are also compared
to the $\omega$ method
which  was introduced in \cite{bb2} as an extension of
the $\Sigma$ method in which  the hadronic final state response is recalibrated
using energy-momentum conservation, and assuming that the hadronic 
miscalibration effects propagate in a similar way in $p_{T,h}$ and $\SSS$,
i.e $\delta p_{T,h} / p_{T,h} \simeq \delta \SSS / \SSS$. This approximation
is correct at $Q^2$ above a few thousands GeV$^2$.
For instance, the resolution of
the inclusive invariant mass M$_{\omega}$, which is related to $x_{\omega}$ 
via M=$\sqrt{x \cdot s}$,
was shown to be excellent at high $Q^2$.  It can be seen in fig.~\ref{newfig8},
how this method is superior to the others at high $x$, while staying comparable
to the best one at low $x$. Using $Q^2_e$ instead of  $Q^2_{\omega}$
in the $\omega$ method (i.e. an ``$e\omega$'' method, not shown) further
improves its purity by about 5 \% in all bins, rendering it optimal in all
these high $Q^2$ bins.

More generally, 
it can be noticed that the stability and purity are significantly
higher at high $Q^2$ 
than at low $Q^2$, thus a finer binning both in $x$ and $Q^2$ 
could be used provided the recorded  number of events is large enough.
In this respect,
the high luminosity HERA upgrade is indeed a very challenging program 
and will allow 
to explore in fine detail the quark-gluon structure of the nucleon
at the highest $Q^2$. These studies will also make use of the
charged current events, for which the kinematics, which has not been studied
here, also improves at high $Q^2$ (see for instance~\cite{gbwh}).
This is due to the improved precision on $p_{T,h}$,
itself coming from the stronger collimation and higher energy of the jet(s).

\section{Conclusion}
After a brief sketch of some experimental
problems  encountered in extracting the structure function
$F_2$ from the measured cross-sections,  
we have reviewed and compared in detail the different ways of
 reconstructing  at HERA the kinematic variables of the inclusive deep 
inelastic scattering. At low $x$,
the effect of rescaling the hadronic energy using both hadronic and
electron observables has been studied and generalized to all methods,
allowing one to derive new properties intermediate between the
$e$ and the $\SSS$ methods. The error propagations 
allow one to choose the most sensible method depending on the kinematic region
under study and on the available precision
of the energy and hadronic calibration. 
The PT method has been compared to the $e$, $\SSS$, 
$r\SSS$ and $e\SSS$ method and is found to be well suited for low $y$ 
but less precise than the other methods at high $y$ (i.e. low $x$) if
the electron energy reconstruction is precise ($\delta$E/E $\simeq$ 1-2\%). In this
case the $e\SSS$ is the most precise method over the full kinematic
range. At high $Q^2$ it has been shown that  the kinematic
reconstruction becomes very precise and that further optimization is
obtained  by using the complete information coming from the electron and
from the hadronic system, as it is done for instance in the $\omega$ method. 
This potential precision underlines
the necessity to have high statistics to reach a systematic
limitation on the measurement and renders
the exploration of the dynamics of
the high $Q^2$ collisions, which will be
done  with the HERA luminosity upgrade program, particularly exciting.

\section{Acknowledgments}
This work has taken place within the
H1 collaboration, and thus has benefitted from the common efforts of all 
our colleagues and we would like to thank them warmly.
In particular we thank Beatriz Gonzalez-Pineiro for the common work we did 
in the 1994 structure function analysis in which we started the studies
introduced in section~2.
Our ZEUS colleagues involved in the structure function measurements
are also thanked for their friendly competition.
We also would like to thank  John Dainton, Robin Devenish, Ralph Eichler, 
and J\"org Gayler for a
careful reading of the manuscript and for their useful remarks.

%---------------------------------------------------------------------- 

\newpage
\noindent
\begin{Large}      
{\bf Appendix:}
% \hspace*{1.5cm}    {\large $y$ and $Q^2$  formulae}   
\end{Large}

%\begin{center}
%Using 
%$\Sigma \equiv \Sigma_h (E_h-p_{z,h}) $, $\ \sss \equiv $E$~(1-\cos{\theta})$ 
%and $r \equiv$ {\Large$ \frac{2 E_{\circ}} {\SSS+\sss}$} we have:
%\end{center}

\begin{table}[h]
\begin{center}
\begin{tabular}{|c|c|c|}  
\hline  
method &  $y$ & $Q^2$ \hhtab\\
%& $x$  \\      
\hline  
 $ e $ &  1 {\large $-\frac{\sss}{2 E_{\circ}}$} &
{\large $\frac{p_{T,e}^2}{1-y_e}$}    \hhtab \\     
\hline      
 $ h $    &  {\large $\frac{\Sigma}{2 E_{\circ}} $}
          &  {\large $\frac{p_{T,h}^2}{1-y_h }$}   
          \hhtab\\ 
%&  $Q^2/ys $\\  
  DA  &  {\large $\frac{\tan\gamma/2}{\tan\gamma/2+\tan\theta/2} $}        
   &  4 E$^{2}_{\circ}$ \  
   {\large$\frac{\cot\theta/2}{\tan\gamma/2+\tan\theta/2}$}
          \\
%&  $Q^2/ys $\\ 
 $ m  $   &  $ y_h   $      
          &  $  Q^2_e $ 
          \hhtab\\
%&  $  Q^2/ys $\\       
\hline  
 $\Sigma$ &  {\large $\frac{\Sigma}{\Sigma+\sss}$}     
          &  {\large $\frac{p_{T,e}^2}{1-y_{\Sigma}}$ }     
          \hhtab\\
%&  $Q^2/ys $\\       
 D$\SSS$  &  $ y_{D\SSS}=   y_{\SSS}   $      
          &  4 E$^{2}_{\circ} \ $ 
{\large$\frac{\cot\theta/2}{\tan\gamma_{\SSS}/2+\tan\theta/2}$}
          \\
%&  $ x_{D\SSS}=x_m $\\ 
 $e\SSS$  &  $  y_{e\SSS}= r \ y_{\SSS}   $      
          &  $  Q^2_e $ 
          \hhtab\\
%&  $  x_{e\SSS}=x_{\SSS} $\\       
\hline  
$r\Sigma$ &  {\large $\frac{r\Sigma}{r\Sigma+\sss}$}     
          &  {\large $\frac{p_{T,e}^2}{1-y_{r\Sigma}}$ }     
          \hhtab\\
%&  $Q^2/ys $\\       
PT ($\equiv r$D$\SSS$)  &  $  y_{PT}=  \ y_{r\SSS}   $      
          &    4 E$^{2}_{\circ} \ $ 
{\large$\frac{\cot\theta/2}{\tan\gamma_{r\SSS}/2+\tan\theta/2}$}
          \\
%&  $Q^2/ys $\\ 
$re\SSS$  &  $  y_{re\SSS}= r_2 \ y_{r\SSS}   $     
          &  $  Q^2_e $ 
          \hhtab\\
%&  $  Q^2/ys $\\       
\hline  
\end{tabular}
\caption  {\label{tabyq}        
\sl 
$y$ and $Q^2$ formulae for the electron and the main hadronic 
methods discussed in the text,    
using: 
%\hspace*{11.2cm} 
$\Sigma \equiv \Sigma_h (E_h-p_{z,h}) $, $\ \sss \equiv $E$~(1-\cos{\theta})$,
$r \equiv$ {\large$ \frac{2 E_{\circ}} {\SSS+\sss}$}
and $r_2 \equiv$ {\large$ \frac{2 E_{\circ}} {r\SSS+\sss}$}
}
\end{center}
\end{table}

\begin{table}
\centering
\begin{tabular}{|c|c|c|c|}
\hline 
        & $h$ &  DA  & \hspace*{0.7cm}   $m$ \hspace*{0.7cm}  \hhtab\\
\hline
        \dyy  &
        \dss  &
{\large   $\frac{1-y_{DA}}{\sin{\gamma}}$} \ $ \delta \gamma - $
{\large   $\frac{1-y_{DA}}{\sin{\theta}}$} \ $ \delta \theta   $ & 
        \dss \hhtab\\
\hline
        \dqq  &
        2 \ \dtt + {\large$\frac{y_h}{1-y_h}$} \ \dss &
{\large  $-\frac{y_{DA}}{\sin{\gamma}}$} \ $ \delta \gamma$ 
{\large  $-(\frac{1}{\sin{\theta}} 
+\frac{y_{DA}} {2 \cos^2{\frac{\theta}{2}}
             \tan{\frac{\gamma}{2}}})$} \ $\delta \theta$    &
$\frac{\delta E}{E}-\tan{\frac{\theta}{2}} \delta \theta$   \hhtab\\
\hline
        \dxx &
        2 \ \dtt + {\large$\frac{2y_h-1}{1-y_h}$} \ \dss &
{\large  $-\frac{1}{\sin{\gamma}}$}  $\  \delta \gamma$
{\large  $-(\frac{y_{DA}}{\sin{\theta}} 
+\frac{y_{DA}} {2 \cos^2{\frac{\theta}{2}}
             \tan{\frac{\gamma}{2}}}$)} \ $\delta \theta$    &
$\frac{\delta E}{E}-\tan{\frac{\theta}{2}} \delta \theta -$\dss \hhtab\\
\hline
\end{tabular}
\caption  {\label{tab22}        
\sl Errors on $y, Q^2$ and $x$ originating from the  errors 
on  the measured variables $\SSS, p_{T,h}, \gamma, E,$
and $\theta$,
for the $h,$ DA and  $m$ methods. 
%For the DA method the errors originating 
%from an error on the electron polar angle are also given. 
}
\end{table}

\begin{table}[htb]
\centering
\begin{tabular}{|c|c|c|c|c|}
\hline 
      & $e$ & $\SSS$ &  D$\SSS$  &   PT ($\equiv r$D$\SSS$)  \hhtab \\
\hline
      \dyy $\frac{1}{\dth}$ &
{\large $\frac{1}{y_{e}}$} 
       {\large $\frac{y_e-1}{\tan{\frac{\theta}{2}}}$} &
{\large $\frac{y_{\SSS}-1} {\tan{\frac{\theta}{2}}}$} &
{\large $\frac{y_{\SSS}-1} {\tan{\frac{\theta}{2}}}$} &
$ $  {\large $\frac{   (2-y_{\SSS}) \ (y_{r\SSS}-1)}{\tan{\frac{\theta}{2}}}$}
\hhtab \\
\hline
      \dqq $\frac{1}{\dth}$ &
$-\tan{\frac{\theta}{2}}  $ &
$-\tan{\frac{\theta}{2}} +$  
      {\large $\frac{1-y_{\SSS}}{\tan{\frac{\theta}{2}}}$} &
$-\tan{\frac{\theta}{2}} -$  
      {\large $\frac{1-y_{\SSS}}{\tan{\frac{\theta}{2}}}$} &
$-\tan{\frac{\theta}{2}} +$  
$ $   {\large $\frac{(2-y_{\SSS}) \ y_{r\SSS}}{\tan{\frac{\theta}{2}}}$}   
\hhtab \\
\hline
      \dxx $\frac{1}{\dth}$ &
$-\tan{\frac{\theta}{2}} - $
      {\large $\frac{1}{y_{e}}$} 
      {\large $\frac{y_e-1}{\tan{\frac{\theta}{2}}}$} &
$-\tan{\frac{\theta}{2}} + $  
      {\large $\frac{2 \ (1-y_{\SSS})}{\tan{\frac{\theta}{2}}}$} &
$-\tan{\frac{\theta}{2}} $  &
$-\tan{\frac{\theta}{2}} +  $
      {\large $\frac{(2-y_{\SSS})}{\tan{\frac{\theta}{2}}}$}   
\hhtab \\
\hline
\end{tabular}
\caption  {\label{tab5}        
\sl Errors on $y, Q^2$ and $x$ due to the error on the polar angle of the 
electron  for the $e, \SSS, $D$\SSS$ and PT  methods.
In the $e\SSS$  method,  $\frac{\delta y}{y}$
is two times larger than for the  $\SSS$ method, i.e.  $\frac{\delta y}{y} = 
2 \ \frac{y_{\SSS}-1}{\tan{\frac{\theta}{2}}} \ \dth$.
In the $r\SSS$  method:  
$\frac{\delta Q^2}{Q^2}=-\tan{\frac{\theta}{2}}+
\frac{2-y_{r\SSS} (2-y_{\SSS})} 
     {\tan{\frac{\theta}{2}}} \ {\delta \theta}.$
}
\end{table}
%\end{small}

\end{document}